\documentclass[aps,onecolumn,superscriptaddress,amsmath,showpacs,tightenlines]{revtex4-1}

\usepackage{graphicx}
\usepackage{subfigure}
\usepackage{natbib}
\usepackage{epsfig}
\usepackage{amsfonts}
\usepackage{amsthm,amsmath,amssymb}
\usepackage{mathrsfs}

\usepackage{epstopdf}
\usepackage{xcolor}
\usepackage[toc,page,title,titletoc,header]{appendix}

\usepackage[utf8]{inputenc}

\begin{document}

\title{The time crystal phase emerges from the qubit network under unitary random operations}
\author{He Wang}
\affiliation{College of Physics, Jilin University,\\Changchun 130021, China}%
\affiliation{State Key Laboratory of Electroanalytical Chemistry, Changchun Institute of Applied Chemistry,\\Changchun 130021, China.}%
\author{Jin Wang}
\email{jin.wang.1@stonybrook.edu}
\affiliation{Department of Chemistry and of Physics and Astronomy, Stony Brook University, Stony Brook,\\NY 11794-3400, USA}%

\begin{abstract}

In this paper, we report findings of non-stationary behavior observed in a fully connected qubit network, utilizing a random unitary evolution model in open quantum system theory. The environmental effect is reflected in the partial swap (PSW) interaction between pairs of qubits with a certain probability. Our study begins with a simple Ising-type Hamiltonian and through many iterations of random unitary evolution, a non-stationary oscillatory state may arise, which encodes certain memory of the initial state. The non-trivial periodic motion of some local observables is indicative of a continuous time crystal phase. We also explore the extension of our study to other types of Hamiltonians and demonstrate that this non-stationary behavior is widespread in our model due to the generalized dynamical symmetry. Remarkably, both theoretical and numerical analysis support the robustness of the constructed time crystal phase to most types of noise.  Our research provides a new perspective for constructing the time crystal phase in an open system model.

\end{abstract}

\maketitle
\section{Introduction}

Understanding the relaxation of the system to the stationary state is a fundamental issue in both classical and quantum statistical mechanics. The former can be accounted for by chaotic dynamics and ergodicity, whereas the latter is more subtle. In accordance with the eigenstate thermalization hypothesis (ETH), relaxation to a stationary state in quantum systems occurs due to the eigenstate dephasing \cite{JD18,MR08}. When there are conserved quantities present, observables relax to stationary values that can be predicted using a generalized Gibbs ensemble distribution \cite{MR2007,EI2015,LV2016}. In parallel with the relaxation to the stationarity at the final, the non-stationary behavior is likewise ubiquitous in Nature, ranging from climate evolution to ecosystem and financial systems, etc. These systems are constantly changing because of external incentives. In recent years, non-stationary behavior in quantum many-body systems has become more prevalent. For example, many-body scarred systems violate strong ETH but still obey weak ETH, exhibiting novel revival dynamics due to the extensive number of non-thermal eigenstates in their spectrum \cite{SMB22}.  Another significant family of systems that exhibit non-stationary motion is the time crystal, which we are particularly interested in. 

In analogy with the common crystal originating from space translation symmetry broken spontaneously, in 2012, Wilczek conceive that time-translational symmetry can also be spontaneously broken, leading to the time crystal \cite{Wilczek12}. The idea was quickly met with some pushback \cite{PN12,PB13}, culminating in the no-go theorem eliminating the possibility of the continuous time crystal (CTC) in the Hamiltonian system with short-range interaction \cite{HM15}. It was soon realized that discrete time crystals (DTC) can exist extensively in the non-equilibrium periodically driven systems, which attract attention both theoretically \cite{KS2015, VK2016, DVE2016, NY2017,NY2020} and experimentally \cite{JZNY17,SCNY17}. Here the system in this case features observables whose expectations break the discrete time-translational symmetry imposed by the external drive.

Essentially, an ideal isolated system does actually not exist, the interaction between the system and the external environment may break down the time crystal phase eventually. For example, the discrete-time crystal appearing in a disordered one-dimensional Ising spin chain cannot endure the coupling to an environment \cite{AL2017}. This is also consistent with down-to-earth observations in the experiments \cite{JZNY17,SCNY17}. The stationarity is the sole ultimate of the system seemingly if it couples with the environment. Nevertheless, it has long been recognized that an open system may contain a decoherence-free subspace in which the states exist unaffected by their surroundings \cite{AB00,PK00}. If the system begins with a state overlapping with the decoherence-free subspace, the system will behave non-stationarily in late time. Appropriately engineered dissipation, also dubbed as quantum-reservoir engineering, can prepare many-body states and non-equilibrium quantum phases, and even perform quantum computation \cite{MBP99,BK08,FV09}. Hence this offers the opportunity to customize the non-stationary state of our interest, i.e., the time crystal phase. There are studies focusing on the DTC in the open system in \cite{ZG18,FMG19,KC22} or beyond \cite{AL20,CMD20,ARC20,RJL22} the mean-field framework, and the experimental observations \cite{JOS20,HK21,HT22}. Studies show that various ways can lead to CTC in the open system. With the generalization of the roton softening mechanism of spatial crystalline, a dissipative Dicke model can exhibit both CTC and DTC phases \cite{XN22}.  A strong continuous measurement on the central spin can induce CTC in a spin star model \cite{MK22}.  Time crystalline behavior, more concretely, boundary time crystal (BTC) can appear on the boundary of the system under the action of the collective Lindblad operators in the thermodynamic limit, whereas for the rest of the system, the bulk remains time-translationally invariant \cite{FI18,GP21}. Also, the CTC may emerge if the system has strong dynamical symmetry \cite{BB2019,CB2020,HA2022}. The CTC has also been observed experimentally in a dissipative atom-cavity system \cite{PK22}. 

In this paper, we investigate the non-stationary behavior of a fully connected qubit network model. While much research has been done on spin chains and lattices with short-range interactions, long-range interactions are important in certain physical systems such as spin glasses \cite{GP06}. Such networks are also of topical interest in quantum information science in the form of quantum communication or quantum computation networks \cite{LMD10}, and their classical counterpart already plays a central role in various branches of classical physics and have been explored extensively in recent decades \cite{AR02}. The network is disturbed by the environment. We model the disturbance of the network by the environment as PSW between arbitrary two sites with a certain probability. The system is first studied with a simple Ising Hamiltonian, and non-stationary oscillations of the local observables are found to appear in all-size networks. The clean CTC phase emerges if the system is prepared in certain initial states. This study is then extended to networks with general Hamiltonians, and a similar conclusion is reached. Therefore, our conclusion is generic. We also discuss the robustness of the constructed time crystal phase. Our study shows the constructed time crystal phase is robust to most types of noise. 

We organize the remainder of the paper as follows. In Sec.\ref{Model and methods}, we introduce our model and the methodology employed. In Sec.\ref{Results}, we outline the key findings of this study. Finally, we draw a conclusion in Sec.\ref{Conclusion}.

\section{Model and methods}\label{Model and methods}

First, we describe the model that interests us, and then we introduce the methods we use to study it.  Following the \cite{BB2019}, the CTC phase of an open system should be defined as a many-body quantum system coupled to a noise-inducing environment that self-organizes in a time-periodic pattern with a period in some observable at the late time for generic initial conditions. For a local observable $ \hat{\pmb O}$, $O(t)=Tr(\pmb\rho(t) \hat{\pmb O})= O(t+T)$ at the  late time, where the continuous-time translation symmetry is spontaneously broken. 

\begin{figure}[!ht]
    \centering
\includegraphics[width=2.5in]{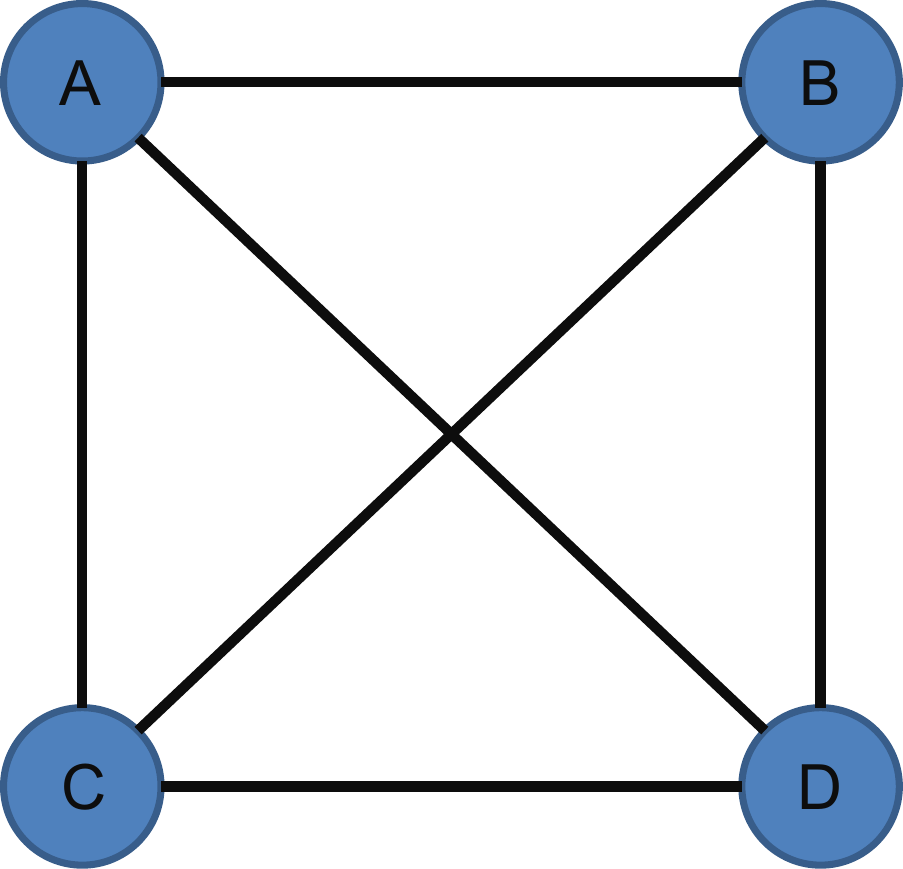}
\caption{\label{fig1} A sketch of the 4-qubit fully connected network.}
\end{figure}

To qualify as a time crystal, the model must be robust against many-body interactions. We consider an all-to-all interacting qubit network system. Its state lives $2^N-d$ Hilbert space $\mathscr{H}$, where $N$ is the number of qubits composed of the network. The linear operators acting on Hilbert space $\mathscr{H}$ define another Hilbert space-- the operator Hilbert space $OP(\mathscr{H})$-- equipped with a Hilbert-Schmidt inner product $(\pmb A, \pmb B) = Tr(\pmb A^\dag \pmb B)$ for all $\pmb A, \pmb B \in OP(\mathscr{H})$. The general form of the Hamiltonian for the fully connected qubit network can be expressed as follows

\begin{equation}\label{HoN}
\pmb H=\sum_{\langle m,n\rangle}J_x\pmb\sigma_x^m \pmb\sigma_x^n+ J_y\pmb\sigma_y^m \pmb\sigma_y^n+ J_z\pmb\sigma_z^m \pmb\sigma_z^n +\sum_{m} h\pmb\sigma_z^m + \sum_{n} t\pmb\sigma_x^n ,
\end{equation}

where $\pmb\sigma_{x}^m (\pmb\sigma_{y}^m, \pmb\sigma_{z}^m)$ are the Pauli matrixes along the x (y,z)-direction for the m-th qubit. We only consider two-body interaction here. The schematic diagram is shown in Fig.\ref{fig1}. It is worth noting that a similar all-to-all interacting qubits Hamiltonian has been used to study boundary time crystals \cite{GP21} and quantum phase transitions \cite{ZBS16}.

We aim to investigate time crystals in an open system, as we mentioned. This is because it is widely acknowledged that no system can truly be considered closed \cite{BH06}, and the environment plays a crucial role in the behavior of most of the systems. To mimic environmental effects, we can use various methods, such as the heat reservoir model \cite{HW22,HW23}, collision model \cite{Xiaoman18,FC22,YL22}, and the random unitary evolution model \cite{JN09,JN10,JN11,JK2015}, which is directly relevant to this paper. Assuming that the initial state of the entire system (system plus environment) is a product state, the random unitary quantum operations can describe the evolution of the system. These operations belong to the class of trace-preserving unital quantum channels and can be expressed as \cite{JN09,JN10,JN11,JK2015}.

\begin{equation}\label{evolution}
\pmb \Phi(\pmb \rho)=\sum_{ m\neq n} p_{mn}\hat{\pmb U}_{mn} \pmb \rho\hat{\pmb U}_{mn}^{\dag}+ p_0 \hat{\pmb U}_{0} \pmb \rho\hat{\pmb U_0}^{\dag} 
\end{equation}

where $\hat{\pmb U}_{mn}=e^{i(\pmb H+\pmb H_{mn})\Delta t}$ and $\hat{\pmb U}_{0}=e^{i\pmb H\Delta t}$ are a set of unitary operators acting on the operator Hilbert space $OP(\mathscr{H})$ of the qubit network. The probabilities of the realizations are denoted by $p_{mn}>0$ and $p_{0}>0$, subject to the constraint that $\sum_{m\neq n} p_{mn} +p_{0} =1$. Here, $\pmb H_{mn}$ describes the interaction or "collision" between the qubits $m$ and $n$, with the specific form of $\pmb H_{mn}$ to be determined later. The parameter $p_{0}$ represents the probability of the system undergoing free evolution, while $p_{mn}$ describes the probability of a collision occurring between the qubit pair $m$ and $n$ during a time interval $\Delta t$, in addition to free evolution. The uncertainty in these probabilities may arise from an unknown error mechanism or a lack of knowledge about subsequent collisions between two sites. For simplicity, we assume a uniform time interval $\Delta t=1$ in the following analysis. After this simplification, the state of the system following n-step evolution is

\begin{equation}
\pmb \rho(n)=\pmb \Phi^n(\pmb \rho(0)).
\end{equation}

In the following discussion, our focus is on the asymptotic quantum states $\pmb \rho(n+1)$ obtained from an initial state $\pmb \rho(0)$ after a large number of iterations n. 

The random unitary operation in Eqn.\ref{evolution} is generally not diagonalizable, which poses challenges in solving the asymptotic dynamics. However, Jaroslav et al. demonstrated that the operator Hilbert space $op(\mathscr{H})$ can be decomposed into a direct sum, namely $OP(\mathscr{H})= Atr(\pmb \Phi) \oplus Atr(\pmb \Phi)^\perp$ \cite{JN10}. Here, $Atr(\pmb \Phi)$ refers to the attractor subspace and $Atr(\pmb \Phi)^\perp$ denotes the orthogonal complement subspace of the attractor subspace \cite{JN10}. The eigenstates corresponding to the eigenvalues of the random unitary operation in Eqn.\ref{evolution} with magnitude $|\nu|=1$ reside in the attractor subspace, while the remaining eigenstates corresponding to the eigenvalues $|\nu|<1$ are located in the orthogonal complement subspace. By transforming the random unitary operation into its Jordan canonical form, one can verify that the components of the orthogonal complement subspace in the density matrix vanish after a sufficiently large number of iterations $\pmb \Phi$. This implies that the asymptotic dynamics of the system, governed by the iterations, are solely determined by the attractor subspace. Moreover, all eigenvectors $\pmb\Gamma$s within the attractor subspace are mutually orthogonal and form a complete basis for the subspace. Therefore, the asymptotic dynamics of the system can be described as

\begin{equation}\label{rho_n}
\begin{split}  
\pmb \rho(n)&=\pmb \Phi^n(\pmb \rho(0))\\
&=\sum_{|\nu|=1}\sum_\alpha^{d_{\nu}} \nu^n \lambda_{\nu,\alpha} \pmb \Gamma_{\nu,\alpha},
\end{split}
\end{equation}

where $ \lambda_{\nu,\alpha}= (\pmb \Gamma_{\nu,\alpha}, \pmb \rho(0)) $ stores the information about the initial state, and $d_{\nu}$ represents the number of eigenvalue $\nu$ in the subspace. Our objective is to determine the attractor subspace, and a useful perspective for this is through the von Neumann entropy $S(\pmb \rho)$, which never decreases under random unitary operations, reads

\begin{equation}\begin{split}  
S(\Phi(\pmb \rho))&=S(\sum_{ m\neq n} p_{mn}\hat{\pmb U}_{mn} \pmb \rho\hat{\pmb U}_{mn}^{\dag}+ p_0 \hat{\pmb U}_{0} \pmb \rho\hat{\pmb U_0}^{\dag})\\
&\geq \sum_{ m\neq n} p_{mn} S(\hat{\pmb U}_{mn} \pmb \rho\hat{\pmb U}_{mn}^{\dag}) + p_0 S(\hat{\pmb U}_{0} \pmb \rho\hat{\pmb U_0}^{\dag})\\
&=S(\pmb \rho).
\end{split}
\end{equation}

The fact that the von Neumann entropy is concave and that its value is unchanged by unitary transformations \cite{EW20} suggests that, for finite-dimensional quantum systems, the entropy tends to be constant in the limit of many iterations. This, along with the monotonicity and boundedness of the entropy, implies that the basis in the attractor space must satisfy the relation simultaneously
\begin{equation}\label{attractor_sol}
\hat{\pmb U}_{mn} \pmb\Gamma_{\nu,\alpha} \hat{\pmb U}_{mn}^{\dag}=\hat{\pmb U}_{0} \pmb \Gamma_{\nu,\alpha}\hat{\pmb U_0}^{\dag}= \nu\pmb \Gamma_{\nu,\alpha},
\end{equation}

for $ \forall m\neq n$. Theorem 4.1 in \cite{JN10} gives a more mathematically rigorous proof. Using Eqn.(\ref{attractor_sol}), we can derive all the eigenvalues in the attractor subspace. Another important insight from Eqn.(\ref{attractor_sol}) is that the asymptotic dynamics are not strongly dependent on the specific probability distribution $p_{mn}$ and $p_0$. Indeed, numerical results suggest that the probability distribution mainly affects the convergence rate towards the asymptotic dynamics \cite{JK2015}. In this paper, we assume $p_0=0.2$ and a uniform $p_{mn}$.  We will perform concrete calculations to find all eigenvectors of the attractor subspace with a simple Hamiltonian in the next section. However, solving Eqn.(\ref{attractor_sol}) for a general many-body Hamiltonian is still a challenging task.

\section{Results}\label{Results}

We have not yet provided the concrete form of the random interaction, $\pmb H_{ij},$ until now. In the following, we specify it as:

\begin{equation}
\pmb H_{mn}=\kappa_{mn}\pmb{SW}_{mn}=\frac{\kappa_{mn}}{2}(\pmb I^m\otimes\pmb I^n+\pmb \sigma_x^m\otimes\pmb \sigma_x^n+\pmb \sigma_y^m\otimes\pmb \sigma_y^n+\pmb \sigma_z^m\otimes\pmb \sigma_z^n),
\end{equation}

with $\pmb{SW}_{mn}=\frac{1}{2}(\pmb I^m\otimes\pmb I^n+\pmb \sigma_x^m\otimes\pmb \sigma_x^n+\pmb \sigma_y^m\otimes\pmb \sigma_y^n+\pmb \sigma_z^m\otimes\pmb \sigma_z^n)$ termed as the swap interaction for the qubits m and n. Its corresponding evolution operator reads

\begin{equation}
e^{i\pmb H_{mn}}=\cos(\kappa_{mn})\pmb I_{mn} + \sin(\kappa_{mn})\pmb{SW}_{mn}.
\end{equation}

This is exactly the partial swap (PSW) operation, which had been used to study the formation of the equilibrium in the dilute quantum gas \cite{JK2015}, the non-Markovity in the collision model \cite{YL22}, quantum thermodynamic engines \cite{MS21}. To simplify matters, we set all $\kappa_{mn}=1$ to avoid PSW degenerating to the trivial case.
Thanks to the full connectivity of the qubit network, it is verified that $[\pmb H_{mn}, \pmb H]=0$. We prove it in Appendix.\ref{appendix.a}. Then Eqn.(\ref{attractor_sol}) decouples into

\begin{equation}\label{attractor_sol_decouple}
\hat{\pmb U}_{0} \pmb\Gamma_{\nu,\alpha} \hat{\pmb U}_{0}^{\dag}=\nu\pmb\Gamma_{\nu,\alpha}, \ \ \pmb{PSW}_{mn} \pmb \Gamma_{\nu,\alpha} \pmb{PSW}_{mn} = \pmb \Gamma_{\nu,\alpha}, \forall m\neq n
\end{equation}

The Eqn.(\ref{attractor_sol_decouple}) shows that the free evolution and the abrupt qubit-qubit PSW interaction dominate each attractor eigenstate. By using the second equation, we can derive the eigenvectors, and subsequently obtain the eigenvalue using the first equation. In the following, we will first start with a simple Hamiltonian to demonstrate the generic non-stationary periodic behavior in some observables and then extend it to more complex Hamiltonians.

To identify the time crystal phase, we monitor the expectation value of local observables $\langle \sigma_x^m \rangle$. Additionally, we can use the Loschmidt echo $LE=Tr(\pmb \rho(0)^{\dag}\pmb \rho(n) )$ as another probe. This has been utilized to differentiate the time crystal phase \cite{CB2020}, as well as in other fields \cite{TM16,FM19,LB21}.

\subsection{A simple Hamiltonian}

We will first consider a simplified Hamiltonian of Eqn.(\ref{HoN}) here. Specifically, we consider a quantum Ising-type Hamiltonian defined as follows:

\begin{equation}
\pmb H_{Ising}=J_z\sum_{\langle m,n\rangle} \pmb\sigma_z^m \pmb\sigma_z^n + h\sum_{m} \pmb\sigma_z^m.
\end{equation}
This Hamiltonian possesses a larger number of conserved quantities due to $[\pmb H_{Ising},\pmb\sigma_z^m]=0$. However, these local conserved charges will be destroyed by random unitary operations. Nonetheless, the total magnetization $\sum_{m} \pmb\sigma_z^m$ remains conserved. Similar one-dimensional qubit chain models have been employed to explore various physical phenomena, such as the many-body localization transition \cite{GY2020}, quantum phase transition \cite{JZ2009}, and quantum computing \cite{MDB2005}.

\begin{figure}[!ht]
    \centering
\includegraphics[width=6in]{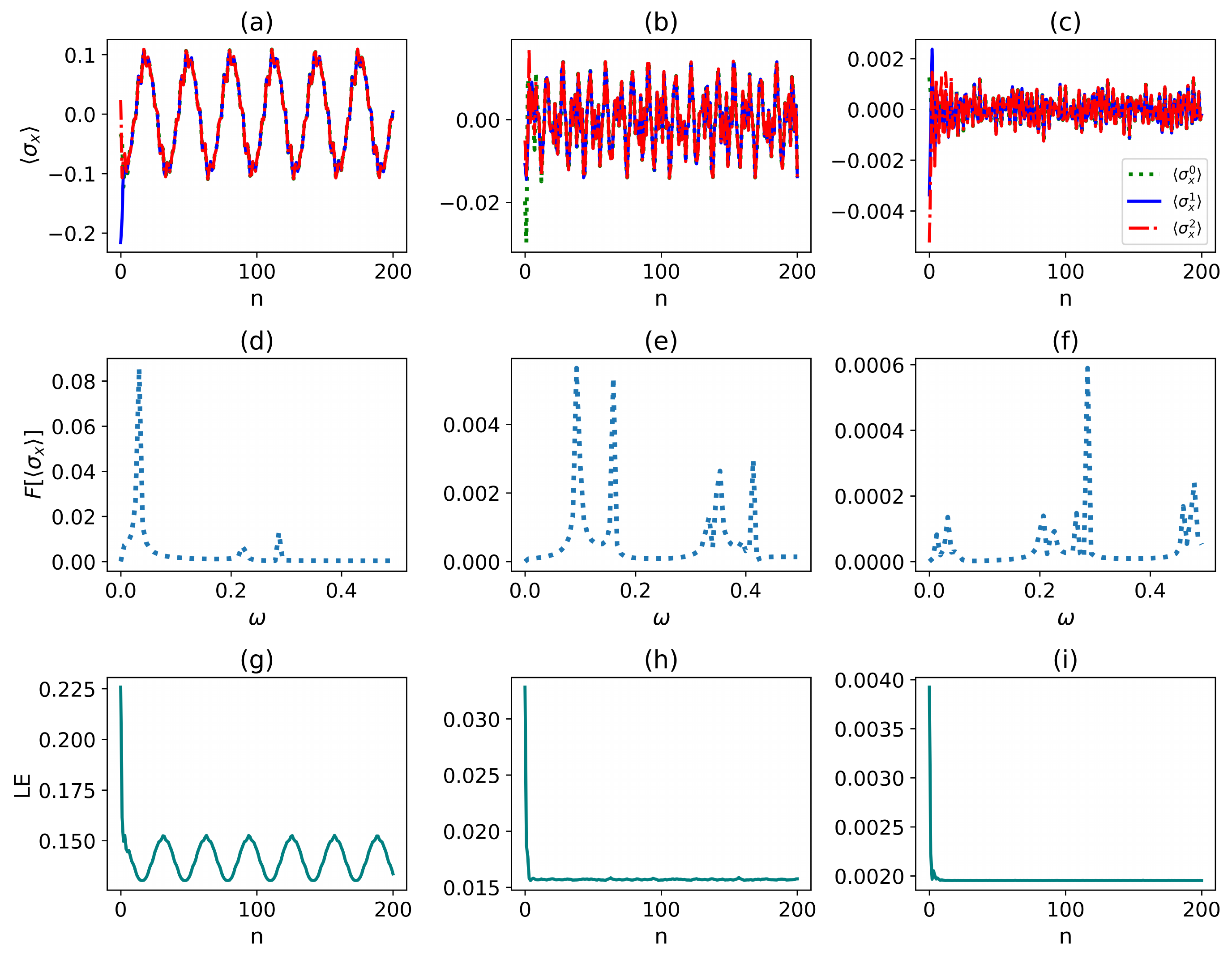}
\caption{\label{fig2} The stroboscopic time evolution of transverse spin of the first three qubits in (a) 3-qubit network, (b)6-qubit network, (c) 9-qubit network. The discrete Fourier transform (DFT) of the transverse spin of the first qubit in each network size is presented in (d,e,f), while the Loschmidt echo is presented in (g,h,i). The networks are initialized in a random state and the values of $h=0.1$ and $k=0.4$ are used.}
\end{figure}

We examine the stroboscopic time evolution of the system by Eqn.\ref{evolution} directly. Initially, the system is in a random state. (Because we do not know what the initial state of the system is with better behavior.) After numerous attempts from various initial states, we have made intriguing discoveries, as illustrated in Fig. \ref{fig2}. In all instances, we uncovered that local observables for various sites converge quickly and oscillate in sync, regardless of network size (as shown in Fig. \ref{fig2}(a-c)). This indicates that our system will not reach equilibrium as long as the PSW interaction exists within the network. The oscillation pattern in the 3-qubit network is quite periodic, indicating a significant CTC phase (Fig. \ref{fig2}(a)).  In the larger qubit network, oscillation is somewhat random but will never be stationary (Fig. \ref{fig2}(b,c)). The Fast Fourier Transform reveals that the 3-qubit network has fewer frequency peaks in its oscillation, whereas the larger-scale qubit network has more frequency peaks (Fig. \ref{fig2}(d,e,f)). The oscillation nature of the entire system is measured by the Loschmidt echo. We observed relatively consistent periodic motion in the small-size network after a certain number of iterations, but oscillations almost vanished in the larger network (Fig. \ref{fig2}(g,h,i)).

Upon the above observations, several interesting questions arise. \textbf{why do local observables oscillate synchronously rather than thermalize?} Fig. \ref{fig2}(d,e,f) suggest that the emergence of a time crystal phase is linked to the appearance of fewer dominant frequencies in oscillations.  \textbf{Therefore, could we create a clean time crystal phase in a general-size network by selecting a specific initial state?}  To address these questions, we'd better study Eqn.(\ref{attractor_sol_decouple}) first.

To solve Eqn.(\ref{attractor_sol_decouple}), we first express it in the computational basis $|i_1 i_2 i_3...i_{N-1} i_{N}\rangle$, where $i_m\in{0,1}$ and $\pmb \sigma_z^{m}|0\rangle=|0\rangle$,  $\pmb \sigma_z^{m}|1\rangle=-|1\rangle$.
Since only two indexes are involved in PSW operation, we can simplify the equation by expressing it in a pair of local indices $|i_m i_n\rangle$ and omitting the others. Therefore, the second equation in Eqn (\ref{attractor_sol_decouple}) can be reformulated as 

\begin{equation}
\pmb \Gamma^{(i_m,i_n)}_{(j_m,j_n)}=\pmb \Gamma^{(i_n,i_m)}_{(j_n,j_m)}
\end{equation}

in the local index. This implies that all matrix elements must be equal by a permutation in the local indices, forming an equivalent class that identifies an eigenbasis of the second equation in Eqn (\ref{attractor_sol_decouple}). Since the order of $  
\left(                 
\begin{array}{c}  
    i_m \\  
    j_m \\  
\end{array}
\right)   $ in the full-index representation is unconsidered, only the number of the local indexes is important. Thus, a class can be specified by the array  $\vec{\beta}=\left(\beta_0^0, \beta_0^1,\beta_1^0,\beta_1^1\right)$, where $\beta_i^j$ is the number of the $     
\left(               
  \begin{array}{c}  
    i \\  
    j \\  
  \end{array}
\right)   $ in the full-index representation. The number of classes, which corresponds to the number of eigenbases of the second equation in Eqn(\ref{attractor_sol_decouple}), is $C_{N+3}^{N}$. The matrix elements in the eigenbasis are identical if they are in the corresponding equivalent class; otherwise, they are zeros. The eigenbasis can be expressed as

\begin{equation}\label{eigenvector}
\pmb \Gamma_{\vec{\beta}}=C\sum_{\pmb\pi\in S_N} |\pmb\pi(i)\rangle\langle\pmb\pi(j)|,
\end{equation}

where the summation is over the symmetric group $S_N$, consisting of all permutations $\pmb\pi$ that acting on the full index $i=(i_1,i_2...i_N)$ \cite{JK2015}. And the normalization coefficient is $C=\frac{1}{\sqrt{N!\beta_0^0!\beta_0^1!\beta_1^0!\beta_1^1!}}$.

We have obtained the complete set of eigenbasis. The next step is to determine the eigenvalue in the first equation in Eqn.(\ref{attractor_sol_decouple}). The Hamiltonian $\pmb H_{Ising}$  is diagonal in the computational basis, and we can easily calculate its energy spectrum and eigenvectors. The eigenvectors of $\pmb H_{Ising}$ can also be classified based on the permutation symmetry of the Hamiltonian. The order of eigenvectors does not matter, and the classes are determined by the total magnetization $\langle\sum_{m} \pmb\sigma_z^m\rangle$, which we call the magnetization class. In general, the eigenenergy $\epsilon(i)$ in different classes is distinct.
Now the first equation in Eqn.(\ref{attractor_sol_decouple}) can be solved. Substitute Eqn.(\ref{eigenvector}) into the first equation in Eqn.(\ref{attractor_sol_decouple})

\begin{equation}\label{eigenvalue}
Ce^{i\pmb H_{Ising}} \sum_{\pmb\pi\in S_N} |\pmb\pi(i)\rangle\langle\pmb\pi(j)| e^{-i\pmb H_{Ising}}=Ce^{i[\epsilon(i)-\epsilon(j)] }\sum_{\pmb\pi\in S_N} |\pmb\pi(i)\rangle\langle\pmb\pi(j)|.
\end{equation}
Therefore, we derive the eigenvectors and corresponding eigenvalues in the attractor subspace. From the Eqn.(\ref{eigenvalue}), only the upper index and the lower index of the eigenvectors $\pmb \Gamma_{\vec{\beta}}$ belonging to the same class, then the eigenvalue $\nu=1$, indicating that there are a large number of eigenvectors corresponding to the eigenvalues $|\nu|=1$ with a limit cycle nature.
To address the first question, we need to study the partial trace of the density matrix after many times iterations.
Referring to the Eqn.(\ref{rho_n}), we only need to study the partial trace of the eigenvectors $\pmb \Gamma_{\vec{\beta}}$. To get the reduced density matrix of the (N-1)-qubit subsystem, we trace out the degree of freedom of one particular qubit. That is

\begin{equation}\label{partial_trace}
\begin{split}
Tr_m(e^{i[\epsilon(i)-\epsilon(j)]}\pmb \Gamma_{\nu,\alpha})&=Ce^{i[\epsilon(i)-\epsilon(j)]}Tr_m(\sum_{\pmb\pi\in S_N} |\pmb\pi(i)\rangle\langle\pmb\pi(j)|)\\
&=Ce^{i[\epsilon(i)-\epsilon(j)]}Tr_m(\sum_{i_m=0,1} |i_m\rangle\langle i_m|\otimes \beta_{i_m}^{i_m}\sum_{\pmb\pi\in S_{N-1}} |\pmb\pi(i^{[m]})\rangle\langle\pmb\pi(j^{[m]})|\\
&+\sum_{i_m=0,1} |i_m\rangle\langle \overline{i}_m|\otimes \beta_{i_m}^{\overline{i}_m}\sum_{\pmb\pi\in S_{N-1}} |\pmb\pi(i^{[m]})\rangle\langle\pmb\pi(j^{[m]})|)\\
&=Ce^{i[\epsilon(i)-\epsilon(j)]}\sum_{i_m=0,1}\beta_{i_m}^{i_m}\sum_{\pmb\pi\in S_{N-1}} |\pmb\pi(i^{[m]})\rangle\langle\pmb\pi(j^{[m]})|\\
&=\frac{e^{i[\epsilon(i)-\epsilon(j)]}}{\sqrt{N}}\sum_{i_m=0,1}\sqrt{\beta_{i_m}^{i_m}} \pmb \Gamma_{\vec{\beta}^{[m]}},
\end{split}
\end{equation}

where $\overline{i}_m=1-i_m$, $i^{[m]}$ denotes for the array $(i_1 ,i_2 ,i_3...i_{m-1},i_{m+1},...i_N)$, and $\vec{\beta}^{[m]}$ denotes for the array $\vec{\beta}$ with the elements $\beta_{i_m}^{i_m}$ being changed to be $\beta_{i_m}^{i_m}-1$. One can go on performing partial trace step by step according to Eqn.(\ref{partial_trace}). We see that the partial trace vanishes if and only if all $\beta_{i_m}^{i_m}=0$. In the one-qubit subsystem, there exist certain eigenvectors with non-unit eigenvalues $e^{i[\epsilon(i)-\epsilon(j)]}$, such as $\frac{1}{2\sqrt{3}}\pmb \Gamma_{(2,1,0,0)}$, $\frac{1}{2\sqrt{3}}\pmb \Gamma_{(0,1,0,2)}$, and $\frac{1}{\sqrt{6}}\pmb \Gamma_{(1,1,0,1)}$, along with their corresponding Hermitian conjugates in the 3-qubit network. By considering only the part with a positive phase in Eqn.(\ref{partial_trace}), we obtain three possible frequencies for the local observables of the one-qubit subsystem, which agrees with our observations in Fig.\ref{fig2}(d)  and other realizations. For the 6-qubit network and the 9-qubit network, there are 6 and 9 possible frequencies for the local observables of the one-qubit subsystem, respectively. Furthermore, Eqn.(\ref{partial_trace}) implies that subsystems of the same size exhibit identical dynamics at late times. Thus, we have fully addressed the first question.

\begin{figure}[!ht]
    \centering
\includegraphics[width=6in]{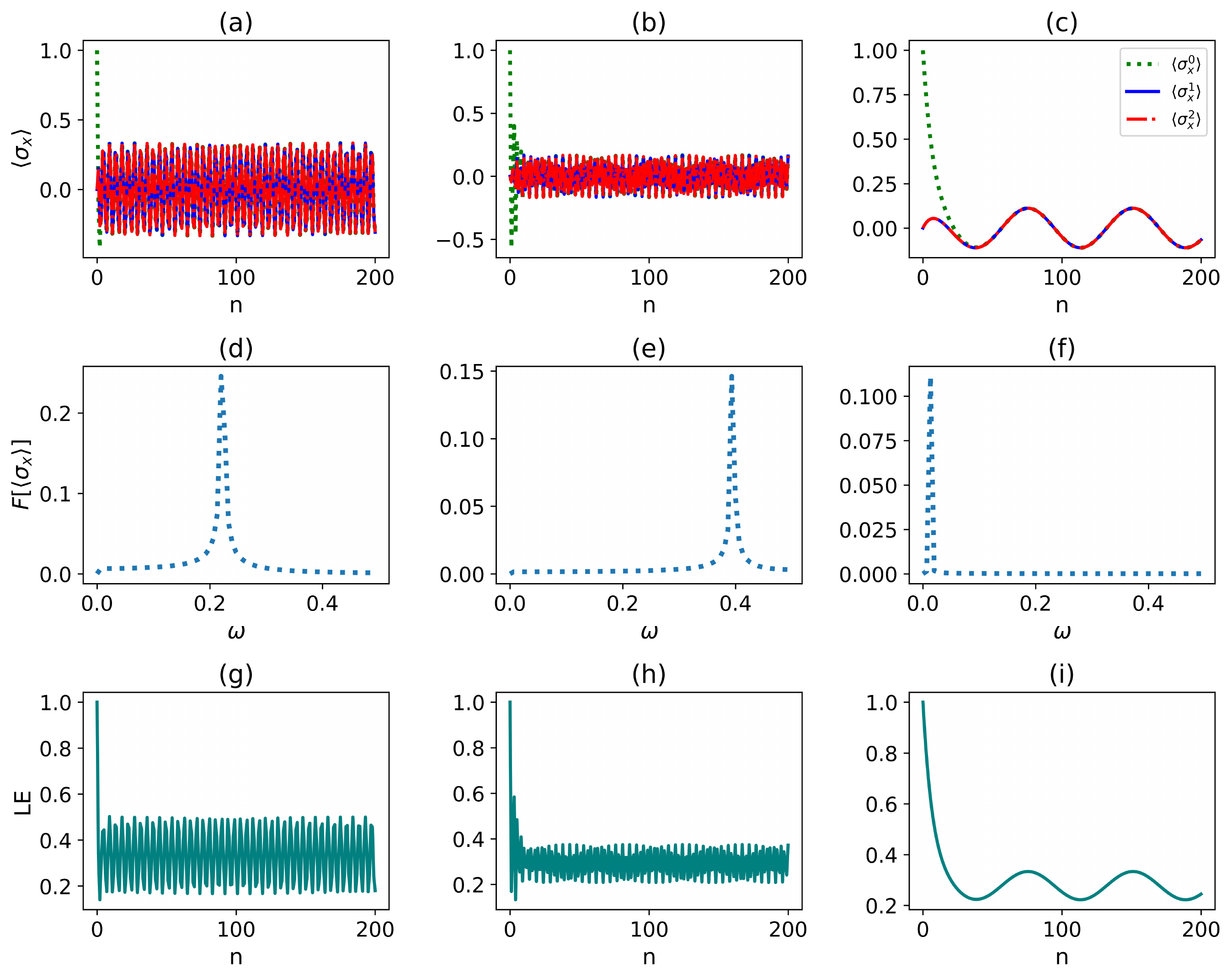}
\caption{\label{fig3} The stroboscopic time evolution of transverse spin of the first three qubits in (a) 3-qubit network, (b)6-qubit network, (c) 9-qubit network. The DFT of the transverse spin of the first qubit is shown in panels (d), (e), and (f) for the three network sizes. The Loschmidt echo for the six network sizes is shown in panels (g), (h), and (i). The networks are initialized in a pure product state, and the values of $h=0.1$ and $k=0.4$ are used.}
\end{figure}

Now we continue to address the second question. To get a clean periodic behavior, we conclude two points from the above analysis. First, the choice of the initial state $\pmb\rho(0)$ is such that $\lambda_{\nu,\alpha}\neq 0$ with $\nu\neq1$, i.e., the initial state has some overlap with the attractor subspace. Second, if there are too many incommensurable phase factors in Eqn.(\ref{eigenvalue}), they will generically dephase as shown in Fig.\ref{fig2}(c).
Here we consider an experimentally accessible initial state. Without loss of generality, the initial pure product state is set as only one qubit at $|+\rangle=\frac{1}{\sqrt{2}}(|0\rangle+|1\rangle)$, the others being at $|0\rangle$. We then examine the evolution of the system, and observe a clean periodic oscillation at late times for all network sizes, as shown in Fig.\ref{fig3}(a,b,c). This behavior is also reflected in the Loschmidt echo in Fig.\ref{fig3}(g,h,i) and in the dominant frequency in the DFT of the transverse spin of the first qubit in Fig.\ref{fig3}(d,e,f). It is important to note that while we chose a specific initial state, there are many other alternative states that can produce similar results. Therefore, the emergence of the time crystal phase is not due to fine-tuning. We conclude this subsection by discussing the time crystal in the thermodynamics limit. In order to understand the behavior of the time crystal phase in the thermodynamic limit, we examine Eqn.(\ref{partial_trace}) and note that all the subsystems of equal size exhibit identical asymptotic behavior, due to the permutation invariance. Considering a local operator $\pmb O_i$ and its corresponding collective operator $\sum_i^N \pmb O_i$, the non-stationary evolution of the expected value of $\pmb O_i$ in the clean time crystal phase can be expressed as $\langle\pmb O_i(n)\rangle =  \frac{1}{N} \sum_i^N\nu^n Tr(\pmb\rho(0)\pmb \Gamma_{\nu}) Tr(\pmb \Gamma_{\nu} \pmb O_i)$. From this expression, we observe that the amplitude of the periodic oscillations of the time crystal phase scales inversely with the size of the system, i.e., it is proportional to $\frac{1}{N}$. To maintain the time crystal phase in the thermodynamic limit, it is necessary to ensure that $Tr(\pmb\rho(0)\pmb \Gamma_{\nu}) Tr(\pmb \Gamma_{\nu} \sum_i^N\pmb O_i)$ is comparable with $N$. This can be achieved by choosing an appropriate initial state for the system. For example, if we prepare the initial state in $\frac{1}{\sqrt{N}}\sum_{i}|00...+_i...00\rangle$ (where $+_i$ denotes that the qubit on the i-th site is in the state $|+\rangle$), the observable $\langle \pmb\sigma_x\rangle$ will exhibit oscillations with the apparent amplitude and the monochromatic frequency, regardless of the system size.

\subsection{Other Hamiltonians}
In the preceding section, we studied the stroboscopic time evolution of the fully connected qubit network with a simple Ising-type Hamiltonian subjected to random unitary operations, which allowed us to realize the time crystal phase in such a system. This leads us to question whether the time crystal phase can be achieved with a more general Hamiltonian. In the following, we investigate the Hamiltonians: transverse-field-Ising (TFI) type, XX type, and XYZ type, as shown below:
\begin{equation}\label{TFI}
\pmb H_{TFI}=\sum_{\langle m,n\rangle}J_z\pmb\sigma_z^m \pmb\sigma_z^n + \sum_{i} t\pmb\sigma_x^n ,
\end{equation}

\begin{equation}\label{XX}
\pmb H_{XX}=\sum_{\langle m,n\rangle}J_x\pmb\sigma_x^m \pmb\sigma_x^n+ J_y\pmb\sigma_y^m \pmb\sigma_y^n+ \sum_{i} h\pmb\sigma_z^n ,
\end{equation}

\begin{equation}\label{XYZ}
\pmb H_{XYZ}=\sum_{\langle m,n\rangle}J_x\pmb\sigma_x^m \pmb\sigma_x^n+ J_y\pmb\sigma_y^m \pmb\sigma_y^n+ J_z\pmb\sigma_z^m \pmb\sigma_z^n +\sum_{i} h\pmb\sigma_z^m,
\end{equation}

 respectively. For the sake of simplicity, we will only consider the 6-qubit network for the remainder of this paper, but our findings will apply to qubit networks of any size.

\begin{figure}[!ht]
    \centering
\includegraphics[width=6in]{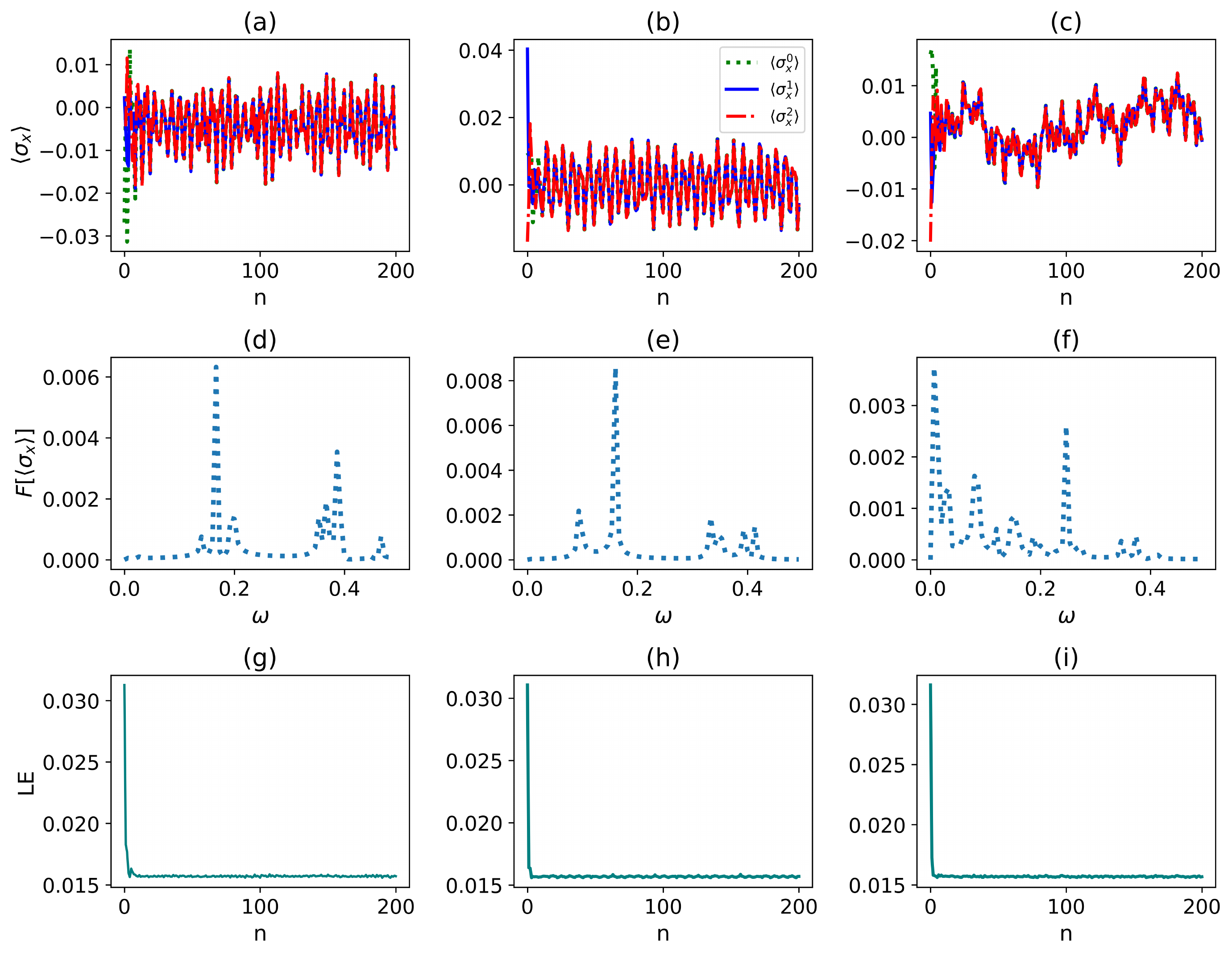}
\caption{\label{fig4} The stroboscopic time evolution of the transverse spin of three qubits in a 3-qubit network with three different types of interactions: (a) TFI-type, with $J_z=0.4$ and $t=0.1$; (b) XX-type, with $J_x=J_y=0.4$ and $h=0.1$; and (c) XYZ-type, with $J_x=0.1$, $J_y=0.2$, $J_z=0.3$, and $h=0.1$. The DFT of the transverse spin of the first qubit with respect to each type of network is presented in panels (d), (e), and (f), respectively. The Loschmidt echo with respect to each type of network is shown in panels (g), (h), and (i), respectively. The network is initialized in a random state.}
\end{figure}

Let us examine the stroboscopic time evolution of the system by Eqn.\ref{evolution} directly as before. Surprisingly, we observe similar phenomena to Fig.\ref{fig2} in Fig.\ref{fig4}. All local observables for different sites converge quickly and oscillate synchronously. They never tend to be stationary in Fig.\ref{fig4}(a,b,c). The DFT spectrum shows that there are certain frequencies in the TFI-type network, with two being significant in Fig.\ref{fig4}(a). There are one and three dominant frequencies in the XX-type and XYZ-type networks, respectively, in Fig.\ref{fig4}(b,c).
The Loschmidt echo will experience a quick drop followed by some permanent minor fluctuations at late times in Fig.\ref{fig4}(g,h,i), implying a time crystal phase. We find similar behavior as in the simple Ising-type network case, even when both begin from a random state. We ask the same questions as in the previous section: \textbf{Why aren't the local observables thermalized and oscillating synchronously?} And \textbf{is it possible to find a clean time crystal phase by choosing a particular initial state?}

To address these questions, we attempt to solve Eqn.\ref{attractor_sol_decouple} as a starting point. The same eigenvectors can be derived in the second equation. However, the first eigen-equation is challenging to solve. An insight from Eqn.\ref{attractor_sol_decouple} is that the first equation with the general Hamiltonian will filter out many eigenvectors in the second equation. Therefore, there aren't as many eigenfrequencies as in the previous case.
We attempt to address the questions by bypassing solving Eqn.\ref{attractor_sol_decouple} directly. We turn back to investigate the eigenvalues and the eigenvectors of the CPTP map Eqn.\ref{evolution}. The eigenvectors related to eigenvalues with the module less than one reside in the orthogonal complement subspace of the attractor subspace $Atr(\pmb \Phi)^\perp$ and cut no ice with the asymptotic dynamics. In the attractor subspace, the eigenvectors with the eigenvalues one stand for the stationary states, and other eigenvectors corresponding to eigenvalues on the unit circle in the complex plane represent the limit cycle dynamics and do not decay. The following theorem yields a straightforward procedure to explicitly establish such asymptotic non-stationary states from the stationary state as well as a set of precise criteria that guarantee their existence in the case of a general quantum channel.

\textbf{Theorem: Consider a completely positive trace-preserving (CTPT) map as Eqn.\ref{evolution} and let $\pmb\rho_{st}$ be one of its stationary states, if the following conditions are satisfied, (i) there exists a system operator $\pmb A$ such that $[\pmb H, \pmb A]=\omega\pmb A $, and (ii) $[\pmb{SW}_{mn},\pmb A]\pmb\rho_{st}=0$ for $\forall m\neq n$, then the operator $\pmb A \pmb\rho_{st}$ evolves according to }\cite{GG22}

\begin{equation}
\pmb \Phi(\pmb A \pmb\rho_{st})= e^{i\omega\Delta t} \pmb A \pmb\rho_{st}, 
\end{equation}

\textbf{with $\omega\in \mathbb{R}$}. The proof can refer to reference \cite{BB2019,GG22}.  Physically, the operator is mentioned as a generalized dynamical symmetry by conditions (i) and (ii) \cite{BB2019,CB2020,GG22,BB22,MM20,BB20MBA}.  A dynamical symmetry of the system's autonomous evolution is defined in particular by condition (i) and condition (ii) requires that this symmetry is insensitive to the random unitary operations. With this in mind, we can now address the first question posed at the beginning of this section. The theorem tells us that, assuming the existence of a generalized dynamical symmetry (which we will explain how to find later), the expectation values of the local observables after many iterations are given by

\begin{equation}
\langle\pmb O(n)\rangle = \sum_i r_i Tr(\pmb O\pmb\rho_{st}^{i})+ \sum_{ij}e^{i\omega n\Delta t}R_{ij} Tr(\pmb O\pmb A_j\pmb\rho_{st}^{i}), 
\end{equation}

with $r_i=Tr(\pmb \rho(0)\pmb\rho_{st}^{i})$ and $R_{ij}=Tr(\pmb \rho(0)\pmb A_j\pmb\rho_{st}^{i})$. Once $Tr(\pmb O\pmb A_j\pmb\rho_{st}^{i})\neq 0$ and $R_{ij}=Tr(\pmb \rho(0)\pmb A_j\pmb\rho_{st}^{i})$ for some $i$ and $j$, then $\langle\pmb O(n)\rangle $ shows a oscillation nature.  As a solution of Eqn.\ref{attractor_sol_decouple}, it must inherit the property of partial trace Eqn.(\ref{partial_trace}). Therefore, subsystems of the same size exhibit the same dynamics at late times. This also explains why all the qubits oscillate synchronously. We have now fully answered the first question. 

The above theorem explains how to construct non-stationary states from stationary states, allowing us to find all the eigenvectors of the attractor subspace based on the stationary states and the generalized dynamical symmetry. However, this is not straightforward. Due to the map Eqn.\ref{evolution} being unital, the maximum mixed state is apparently a stationary state. To find other stationary states, a low-efficiency strategy is to initialize a random state and let it evolve, judging the final state as stationary if all local observables converge to constants at a late time. Generally, unstable oscillatory behavior is more likely to be observed. In the following, we only consider the maximum mixed-state scenario.

\begin{figure}[!ht]
    \centering
\includegraphics[width=6in]{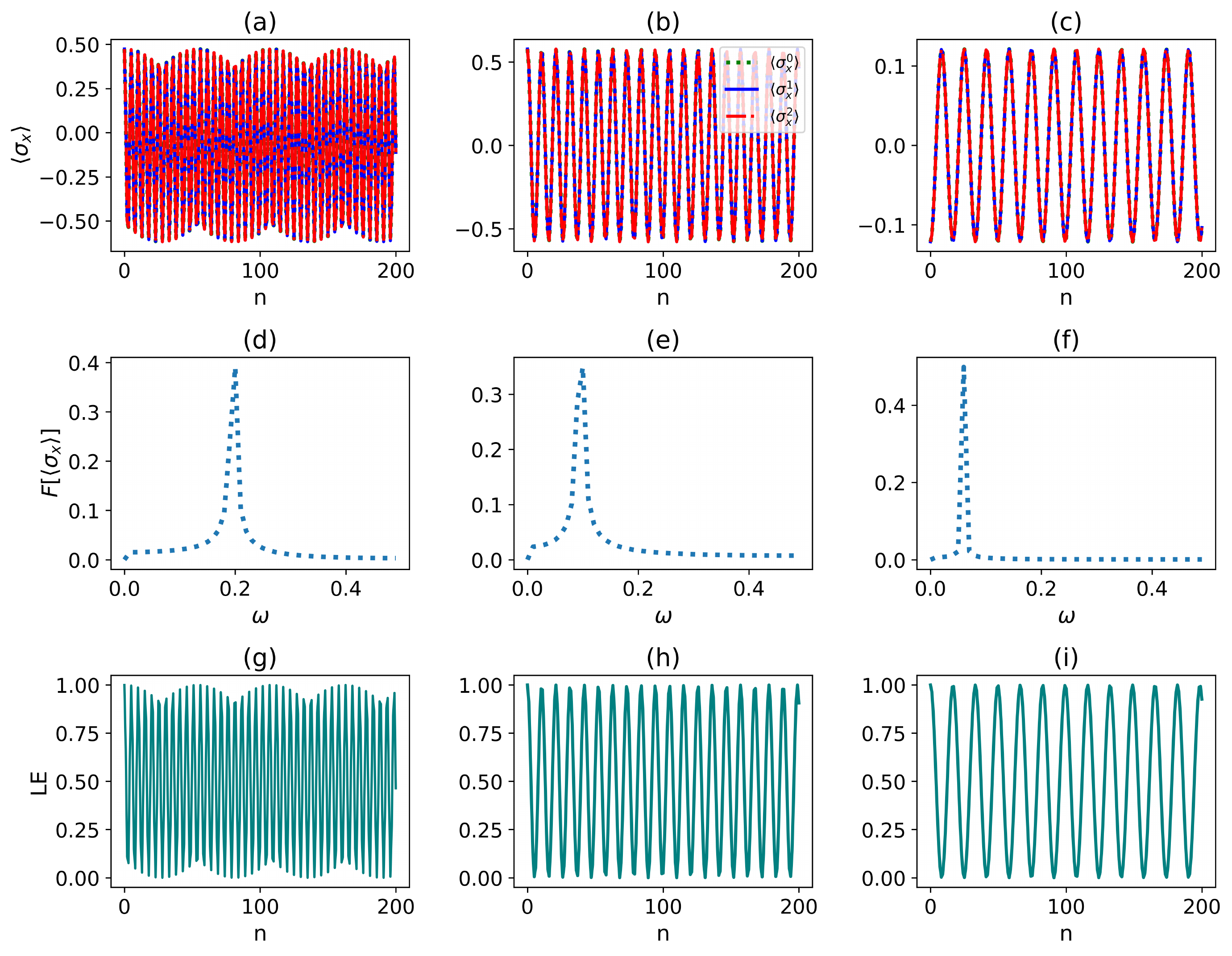}
\caption{\label{fig5} The stroboscopic time evolution of the transverse spin of three qubits in a 6-qubit network, under three different types of interactions: (a) TFI-type with $J_z=0.4$ and $t=0.1$, (b) XX-type with $J_x=J_y=0.4$ and $h=0.1$, and (c) XYZ-type with $J_x=0.1$, $J_y=0.2$, $J_z=0.3$, and $h=0.1$. The DFT of the transverse spin of the first qubit with respect to each type of network is shown in panels (d), (e), and (f), while the Loschmidt echo for each network is displayed in panels (g), (h), and (i). The network is initialized in a specific pure state.}
\end{figure}

We now move to seek generalized dynamical symmetries. Finding generalized dynamical symmetry is quite demanding in the presence of a generic environment and there is even no such symmetry at all. Nevertheless, the generalized dynamical symmetry always exists in our setting. We prove this statement from the beginning of $[\pmb H,\pmb H_{mn}]=0 $ for $\forall m\neq n$. The following steps can be taken: 
(a) Diagonalize one of $\pmb H_{mn}$, rank its eigenvalues (which are only $1$ or $-1$) and the corresponding eigenvectors. (b) Transform the system Hamiltonian $\pmb H$ with a unitary transformation constructed from the ordered eigenvectors of $\pmb H_{mn}$. The system Hamiltonian $\pmb H$ is now on the basis of the eigenvectors of $\pmb H_{mn}$ and is block diagonalized. (c) Diagonalize $\pmb H$, focusing on the eigenvalues and corresponding eigenvectors in a particular block, the block with the eigenvalue of $\pmb H_{mn}$ being $1$, for example. All the eigenvectors of $\pmb H$ in this block can be certain linear superpositions of the eigenvectors with eigenvalues $1$ of $\pmb H_{mn}$. (d) Perform the same procedure on all $\pmb H_{mn}$ and collect all the common eigenvalues in the block as a set. Transform the eigenvectors with eigenvalues in the set back to the computational basis. Any pair of eigenvectors construct a generalized dynamical symmetry.
Now we can address the second question. Following the above steps, we can construct different generalized dynamical symmetries for the various Hamiltonian. For example, $|E_0\rangle\langle E_{49}|$ and its hermitian conjugate for the TFI-type Hamiltonian, $|E_{62}\rangle\langle E_{63}|$ and its hermitian conjugate for the XX-type Hamiltonian, $|E_{61}\rangle\langle E_{62}|$ and its hermitian conjugate for the XYZ-type Hamiltonian, where $|E_i\rangle$ is the eigenvector of one of the three Hamiltonians with eigenvalue $E_i$ in order. To realize the clean time crystal phase, one can start with pure states $1/\sqrt{2}(|E_0\rangle+|E_{49}\rangle)$, $1/\sqrt{2}(|E_{62}\rangle+|E_{63}\rangle)$, $1/\sqrt{2}(|E_{61}\rangle+|E_{62}\rangle)$ for the network of the different type. The result shows in Fig.\ref{fig5}. The local observables oscillate synchronously with a single frequency in the 6-qubit networks in Fig.\ref{fig5}(a,b,c). Their frequencies are related to the initial states in Fig.\ref{fig5}(d,e,f). The LE does the same oscillation motion pattern forever in Fig.\ref{fig5}(g,h,i). Therefore, we completely addressed the second question. It is worth noting that the presence of such a dynamical symmetry is widespread throughout our model. To achieve a clean time crystal phase, one could begin with readily feasible experimental pure states that display overlaps with a variety of eigenvectors. By doing so, the final time crystal phase may contain multiple frequencies.

\subsection{The robustness of the time crystal phase}\label{The robustness of the time crystal phase}

Although we used uniform probabilities $p_{mn}$ in our text, non-uniform probabilities would still lead to the same results, as we have previously mentioned. Furthermore, the powerful theorem guarantees that even if there are some random fluctuations in the swap interaction strength $\kappa_{mn}$, the result remains unchanged (excluding some parameters which make the interaction trivial). The resulting disorder robustness is also guaranteed in time, which means that even if the probability distribution and the strength of the collisions may differ in a single collision, as long as the initial state is properly selected, we can still obtain the clean non-steady oscillatory behavior for $lim_{n\rightarrow\infty}\pmb \Phi_n ...\pmb \Phi_3\pmb \Phi_2 \pmb \Phi_1 (\pmb \rho)$ in the long time limit.

We also investigate the scenario where the condition $[\pmb H_{mn},\pmb H]=0$ is weakly broken. To this end, we introduce classical noise to the system, which results in a non-uniform on-site Hamiltonian of the single qubit and the inter-qubit coupling. This leads to the modification of the system Hamiltonian, which becomes $\pmb H_p=\pmb H + \varepsilon\pmb H' +\mathscr{O}( \varepsilon^2)$. Here, $\varepsilon$ is a small parameter, and $H'$ is a non-uniform Hamiltonian. As a result, $[\pmb H_{mn},\pmb H_p]\neq0$. To analyze the effects of the introduced noise, we take the continuum time limit and keep $\Delta t$ to the first order. In this context, it is helpful to rewrite the system's evolution in the Liouvillian form:

\begin{equation}\begin{split}
\frac{d\pmb \rho}{dt}=\mathscr{L}(\pmb \rho)=-i[\pmb H_p, \pmb \rho]-i\sum_{m\neq n}p_{mn}[\pmb H_{mn},\pmb \rho]+\mathscr{O}(\Delta t^2), 
\end{split}
\end{equation}

The first term represents the unitary evolution of the system, while the second term describes the environmental effect. Under the conditions (i) $[\pmb H, \pmb A]=\omega\pmb A $ and (ii) $[\pmb{SW}_{mn},\pmb A]\pmb\rho_{st}=0$ for $\forall m\neq n$, it can be verified that $\frac{d\pmb A\pmb \rho_{st}}{dt}=-i\omega\pmb A\pmb \rho_{st}$ for the unperturbed system. In the following, we focus on the density matrix $\pmb \rho$ corresponding to the purely imaginary eigenvalues of Liouvillian. To investigate the robustness of the time crystal phase against noise, we split the Liouvillian $\mathscr{L}$ according to the order of the $\varepsilon$,

\begin{equation}\begin{split}
\mathscr{ L}&=\mathscr{ L}^{(0)}+\varepsilon\mathscr{ L}^{(1)},\\
\mathscr{L}^{(0)}&=-i[\pmb H, \cdot]-i\sum_{m\neq n}p_{mn}[\pmb H_{mn},\cdot]\\, 
\mathscr{L}^{(1)}&=-i[\pmb H',\cdot].
\end{split}
\end{equation}

Accordingly, the superket, superbra, and associated eigenvalue can be expanded as 

\begin{equation}\begin{split}
|\pmb\rho\rangle\rangle &=|\pmb\rho^{(0)}\rangle\rangle+\varepsilon|\pmb\rho^{(1)}\rangle\rangle+...,\\
\langle\langle\pmb\eta|&=\langle\langle\pmb\eta^{(0)}|+\varepsilon\langle\langle\pmb\eta^{(1)}|+...,\\
\lambda&=\lambda^{(0)}+\varepsilon\lambda^{(1)}+....
\end{split}
\end{equation}
where $\lambda^{(0)}$ is purely imaginary. To proceed with our analysis, we use the condition $\langle\langle\pmb\eta|\pmb\rho\rangle\rangle=Tr(\pmb\eta^\dagger\pmb\rho)=1$. This yields $Tr(\pmb\eta^{(0)\dagger}\pmb\rho^{(0)})=Tr(\pmb\eta^{(0)\dagger}\pmb\rho^{(1)}+\pmb\eta^{(1)\dagger}\pmb\rho^{(0)})=1$ to the first order. We then expand $\langle\langle\pmb\eta|\mathscr{L}|\pmb\rho\rangle\rangle$ to the first order, which gives:

\begin{equation}
\lambda^{(1)}=\lambda^{(0)}+Tr(\pmb\eta^{(0)\dagger}\mathscr{L}^{(1)}\pmb\rho^{(0)}).
\end{equation}

We can determine that $Tr(\pmb\eta^{(0)\dagger}\mathscr{L}^{(1)}\pmb\rho^{(0)})$ is a real negative number due to the hermiticity of $\pmb\eta$ and $\pmb\rho$. The behavior of the time crystal phase can be described as $\langle\pmb O(t)\rangle=Tr((\pmb\eta^{(0)\dagger}+\varepsilon\pmb\eta^{(1)\dagger})\pmb O)e^{(\lambda^{(0)}+\varepsilon\lambda^{(1)})t}$. Thus, the life of the time crystal phase is proportional to $\mathscr{O}(\varepsilon^{-1})$ in the presence of noise. 

\begin{figure}[!ht]
    \centering
\includegraphics[width=6in]{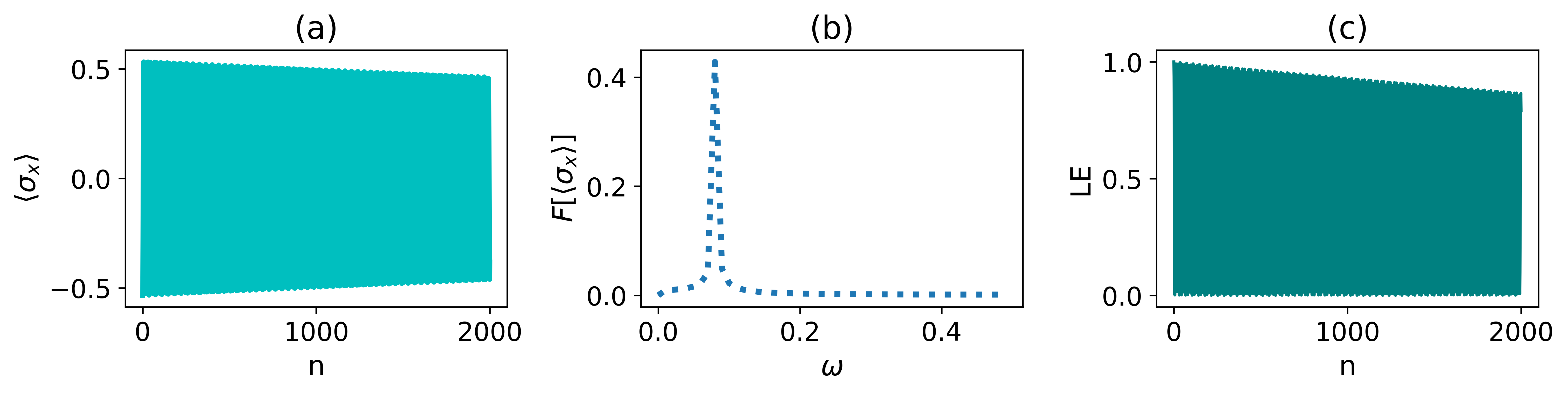}
\caption{\label{fig6} The stroboscopic time evolution of the transverse spin of the first qubit in a 6-qubit network in (a). The Hamiltonian is of XX-type with $J_x^i=J_y^i=0.4+\varepsilon\mathscr{N}(0,1)$ and $h=1+\varepsilon\mathscr{N}(0,1)$, where $\mathscr{N}(\mu,\sigma^2)$ is the Gaussian distribution with the mean $\mu=0$ and the variance $\sigma^2=1$. And $\varepsilon=0.1$. The DFT of the transverse spin of the first qubit is shown in panel (b), while the Loschmidt echo is displayed in panel (c). The network is initialized in a specific pure state $1/\sqrt{2}(|E_{62}\rangle+|E_{63}\rangle)$. }
\end{figure}

We plot the spin dynamics of the XX-type qubit network suffering from the random noise in Fig.\ref{fig6}. We take a relatively larger perturbation value $\varepsilon=0.1$ compared to the energy scale of the system. Despite this, we found that the amplitude of the periodic oscillation of the observable decreases very slowly in Fig.\ref{fig6}(a), as does the Loschmidt echo in Fig.\ref{fig6}(c). The DFT spectrum shows that the system still exhibits oscillations of a single frequency despite the presence of noise in Fig.\ref{fig6}(b). Therefore, the constructed time crystal phase is long-lived under the condition that $[\pmb H_{mn},\pmb H_p]\neq0$ is weakly broken. Taken together, these analyses provide strong evidence that the constructed time crystal phase is robust to most types of noise. This robustness is a promising feature for the potential experimental observations of our model.

\section{Conclusion}\label{Conclusion}

In summary, we have introduced a fully connected qubit network under random unitary operations. The environmental effect
is described as the partial swap occurring on any pair of qubits with probability. We first consider a simple Ising Hamiltonian of the network. We observed non-stationary oscillation motion for local observables at late times with random initial states. Their later dynamics synchronize and the system is not thermalized. This holds for all the scale networks. We explain these phenomena with the attractor subspace theory and then construct a clean time crystal phase for the system. We extend our study to the network with a general Hamiltonian. Three types of Hamiltonians are considered.
Numerical results show that non-stationary behavior is generic for the general Hamiltonians. We can not find all the elements in the attractor subspace due to the complexity of the general Hamiltonians. However, We can construct the elements in the attractor subspace from the stationary states with the help of the generalized dynamical symmetry. And also, we can uncover a clean time crystal phase from certain initial states. In order to assess the feasibility of experimental realizations, it is important to investigate the robustness of the constructed time crystal phase to noise. Our study demonstrates that the time crystal phase constructed in our model is indeed robust to most types of noise. Our study opens a new opportunity to realize the time crystal phase in the open system. Lastly, we point out the feasibility of our model in state-of-the-art experimental capabilities. All the different elements for the construction of the fully connected network are already in place in the laboratory.  The circuit quantum electrodynamics (QED) provides a natural platform in which a large number of qubits can be coupled together \cite{DIS08,SA08,AB21}. Many qubits can be connected together naturally using the cQED architecture. Superconducting qubits serve as the atoms in such systems, and a harmonic oscillator circuit element is in the capacity of a cavity with which they interact. A single cavity will mediate coupling between all possible qubit pairs if it is connected to all qubits at once. If the cavity is also far off resonance with the qubits, its degrees of freedom can be integrated out of the problem, giving us a system with pairwise interactions between every qubit \cite{SA08}. In \cite{XK20}, authors successfully probed the out-of-equilibrium behavior of a spin model in a programmable quantum simulator with 16 all-to-all connected superconducting qubits.  Moreover, numerous effective schemes for implementing quantum gates on superconducting qubits in QED have been proposed \cite{SC18,RY20,ZYZ21}. Therefore, it is possible for our model to be implemented in the experiment.

\appendix

\section{The proof of $[\pmb H_{mn},\pmb H]=0$}\label{appendix.a}

In order to prove $[\pmb H_{mn},\pmb H]=0$, where $\pmb H$ and $\pmb H_{mn}$ are defined as $\pmb H=\sum_{\langle m,n\rangle}J_x\pmb\sigma_x^m \pmb\sigma_x^n+ J_y\pmb\sigma_y^m \pmb\sigma_y^n+ J_z\pmb\sigma_z^m \pmb\sigma_z^n +\sum_{m,j} h_j^m\pmb\sigma_j^m $ and $\pmb H_{mn}=\frac{\kappa_{mn}}{2}(\pmb I^m\otimes\pmb I^n+\pmb \sigma_x^m\otimes\pmb \sigma_x^n+\pmb \sigma_y^m\otimes\pmb \sigma_y^n+\pmb \sigma_z^m\otimes\pmb \sigma_z^n)$. To do this, we begin by deriving the expression for $[\pmb H_{mn},\pmb\sigma_i^m \pmb\sigma_i^n]$.

\begin{equation}\begin{split}
[\pmb H_{mn},\pmb\sigma_i^m \pmb\sigma_i^n]&=\sum_{j=x,y,x}[\pmb\sigma_j^m \pmb\sigma_j^n,\pmb\sigma_i^m \pmb\sigma_i^n] \\
&=\sum_{j=x,y,x}(\pmb\sigma_j^m \pmb\sigma_i^m)\otimes(\pmb\sigma_j^n\pmb\sigma_i^n)-(\pmb\sigma_i^m \pmb\sigma_j^m)\otimes(\pmb\sigma_i^n\pmb\sigma_i^n),
\end{split}
\end{equation}

Combining the commutation relationship $[\pmb\sigma_a,\pmb\sigma_b]=2i\varepsilon_{abc}\pmb\sigma_c$ and anti-commutation relationship $\{\pmb\sigma_a,\pmb\sigma_b\}=2\delta_{ab}$, we obtain $\pmb\sigma_a\pmb\sigma_b=i\varepsilon_{abc}\pmb\sigma_c+\delta_{ab}$. With the help of this relationship, we get
\begin{equation}\begin{split}
[\pmb H_{mn},\pmb\sigma_i^m \pmb\sigma_i^n]&=\sum_{j=x,y,x}(\pmb\sigma_j^m \pmb\sigma_i^m)\otimes(\pmb\sigma_j^n\pmb\sigma_i^n)-(\pmb\sigma_i^m \pmb\sigma_j^m)\otimes(\pmb\sigma_i^n\pmb\sigma_i^n)\\
&=\sum_{j=x,y,x}(i\varepsilon_{jik}\pmb\sigma_k^m+\delta_{ji})\otimes(i\varepsilon_{jik}\pmb\sigma_k^n+\delta_{ji})-(i\varepsilon_{ijk}\pmb\sigma_k^m+\delta_{ij})\otimes(i\varepsilon_{ijk}\pmb\sigma_k^m+\delta_{ij})\\
&=0,
\end{split}
\end{equation}

Next, we examine$ [\pmb H_{mn},\sum_{m,j} h_j^m\pmb\sigma_j^m]$. To simplify the calculation, we can consider an arbitrary pair of $\sum_{m,j} h_j^m\pmb\sigma_j^m$, and we derive

\begin{equation}\begin{split}
[\pmb H_{mn},h_i^m\pmb\sigma_i^m +h_i^n\pmb\sigma_i^n]&=\sum_{j=x,y,x}[\pmb\sigma_j^m \pmb\sigma_j^n,h_i^m\pmb\sigma_i^m +h_i^n\pmb\sigma_i^n] \\
&=\sum_{j=x,y,x}2ih_i^m\varepsilon_{jik}\pmb\sigma_k^m\otimes\pmb\sigma_j^n+2ih_i^n\varepsilon_{ijk}\pmb\sigma_j^m\otimes\pmb\sigma_k^n\\
&=2i(h_i^n-h_i^m)\varepsilon_{ijk}\pmb\sigma_j^m\otimes\pmb\sigma_k^n,
\end{split}
\end{equation}
This implies that $ [\pmb H_{mn},\sum_{m,j} h_j^m\pmb\sigma_j^m]=0$ if and only if $h_i^n-h_i^m=0$, which is exactly our case. Hence we complete the proof of $[\pmb H_{mn},\pmb H]=0$.

\end{document}